\documentclass[aps,prl,twocolumn,preprintnumbers]{revtex4-1}
\usepackage{graphicx,amsmath,amssymb,bm,color,slashed}
\newcommand{\eq}[1]{Eq.~(\ref{#1})}
\newcommand{\be}{\begin{equation}}
\newcommand{\ee}{\end{equation}}
\newcommand{\bea}{\begin{eqnarray}}
\newcommand{\eea}{\end{eqnarray}}

\def\pp{{p'}}
\def\G{{\Gamma}}
\def\S{\Sigma}

	 \def\e{{\epsilon}}

\def\L{{\Lambda}} \def\l{{\lambda}} 
\def\K{{\kappa}}
\def\l{\lambda}
  \def\S{\Sigma}\def\D{\Delta}\def\la{\langle}\def\ra{\rangle}\def\s{\sigma}

\def\m{{\mu}}\def\k{{\kappa}}\def\p{{\phi}}\def\g{\gamma}\def\d{\delta}\def\n{{\nu}}
\def\psl{\slashed{p}}\def\ksl{\slashed{k}}
\def\Ksl{\slashed{\kappa}}
\def\qsl{\slashed{q}}\def\on{{\rm_{on}}}\def\cU{{\cal U}}\def\part{{\partial \over \partial M^2}}
\def\pd{\partial}

\begin{document}

%\catchline{}{}{}

%\preprint{NT@UW-09-18}
\title{\hskip 4.5cm  % letter.tex V0$\quad$ 
NT@UW-19-07\\ Confinement in Nuclei and the Expanding Proton } 

%preprint{NT@UW-09-??}
\author{ %\footnotesize 
Gerald   A. Miller}

\affiliation{Department of Physics,
University of Washington\\
Seattle, Washington 98195-1560
}

\begin{abstract}   High-precision knowledge of electromagnetic form factors of nuclei is  a subject of much current  experimental and theoretical activity in  nuclear and atomic physics.   Such precision mandates that  effects of the non-zero spatial extent of the constituent nucleons be handled in a manner that goes beyond  the usual  impulse  approximation. %in which nucleon electromagnetic form factors are  taken to be the same as in free space.  
A   series of simple, Poincare-invariant, composite-proton models that respect the Ward-Takahashi identity and in which quarks are confined are used to study the validity of this approximation.   The result of all of the models is a general theorem showing   that medium modification of   proton structure must occur.  Combining  this result with  lattice QCD calculations  leads to a conclusion
  that a bound proton must be larger than a free one.    
 
\end{abstract}
\date{\today}
\maketitle

Nucleons are composite particles made of quarks, gluons and quark-pairs bound   by the confining forces of QCD.  The composite nature means that nucleons bound in nuclei must be different than free ones~\cite{Sargsian:2002wc}.   Many years of experiment and theory tells us  the answer: the differences exist but are  not very  large. %Even so, the effects of the medium on the structure of a bound nucleon cannot vanish. 
Early evidence of that was found in the EMC effect \cite{Aubert:1983xm,Arnold:1983mw} and also in kaon-nucleus scattering~\cite{Mardor:1990cr}.   Recent reviews are in Refs.~\cite{Hen:2016kwk,Cloet:2019mql}, and a recent update with many references is~\cite{Wang:2018wfz}.
The present manuscript aims to present a new approach to medium modification of the nucleon wave function that is related both to experiment and to lattice QCD calculations.  The key result of the presented arguments is that the proton gets bigger when it is bound in a nucleus.\\

The specific focus here is on elastic electron-nucleus scattering. This reaction has the simplifying  feature that the initial and final nuclei are in the same quantum state. Elastic electromagnetic form factors of nuclei  can be compared with  {\it ab initio} nuclear structure calculations. For example,   \cite{Ruiz:2016gne}   measures isotope shifts in  the radii of Ca isotopes to better than 1\% accuracy. New muonic atom measurements~\cite{Antognini:2015moa}  that determine the charge radii of light nuclei are now at about the 1\% level.   Furthermore,  a  current  Jefferson Laboratory experiment~\cite{12,Gomez:2017cwj}  aims to measure the  the difference between the charge radii of  $^3$He and $^3$H to a precision of $\pm0.02$ fm.   The  current  high precision goals  create  a need to  learn  how to improve the treatment of  the effects of the non-zero spatial extent of the constituent nucleons.\\ % to better accuracy than previously. \\

This is because
the nuclear electromagnetic form factor ${\cal F}_A(Q^2)$   is  often approximated  as:
\bea{\cal F}_A(Q^2)=F_A(Q^2)G_E(Q^2),  \label{FA}\eea
where a spin-0 nucleus  absorbs a space-like photon of four momentum $q^\m$ and $Q^2=-q^2$,  $G_E(Q^2)=F_1(Q^2)-{Q^2\over 4M^2}F_2(Q^2)$ is the proton Sachs electric form  factor,  where  $F_{1,2}$  are Dirac and Pauli   form factors,  $F_A(Q^2)$ is 
the probability   amplitude for a point proton to absorb momentum without changing the nuclear state and $M$ is the proton mass. 
Effects of other charged particles are ignored here for simplicity.
This product ansatz of \eq{FA}  will be referred to as the {\it factorization approximation}.  \\ %For spin-1/2 nuclei, the form factors $ {\cal F}_A$ and $F_A$ are the non-spin-flip amplitude.\\

The original derivation of  \eq{FA}   is ancient~\cite{Amaldi1950,Villi1959} and is based on non-relativistic classical physics, 
A quantum mechanical derivation of the same formula can  be obtained by assuming an impulse approximation in which only the free form factors,
 $F_{1,2}(Q^2)$  appear.
 %   even when  the initial proton four-momentum $p$  and the final proton momentum $p'\equiv p+q$ are not on the mass-shell ($p^2 (\pp^2)\ne M^2$), see Fig.~1a.  In particular, the  factorization approximation assumes that the relevant amplitudes have no dependence on $p$ or $p'$. 
 The factorization  approximation cannot be 100\%  accurate because the struck protons are bound in nuclei.  The  factorization approximation has been  widely used for a long time, with no examination appearing in the literature. %Modern-day precision demands a study of its accuracy.
  \\

The  approach taken here  is to construct a diverse set of models of the free  proton and then place that proton in the nucleus. Elastic electron-proton scattering is shown in   Fig.~1a.    In free space, $p^2=\pp^2=M^2$.  The initial and final protons are on their mass shell. Imposing parity, time-reversal invariance, Lorentz symmetries and current conservation the observable quantities are the Dirac (Pauli) form factor $F_{1,(2)}(Q^2)$. %, where $Q^2$ is negative of the space-like square of the virtual-photon momentum.
 Suppose  instead the 
 proton is bound in the nucleus (see Fig.~1b).    Interactions with nuclei involve evaluating Feynman graphs containing  an integral over the four-momentum $p$ of the initial nucleon that  ranges over all possible values of $p^2$  from $-\infty$ to $\infty$.  % In Fig.~1a a proton of momentum $p$ consists of a quark  of momentum $k$ and a spectator system  of momentum $p-k$. The quark absorbs a photon of momentum $q$ and has a probability amplitude to be part of a final proton of momentum $p+q$.
 This means that  the  Einstein relation equality between square of the four-momentum and $M^2$  is not generally accurate for bound nucleons.  In general, the nucleon form factors should depend on $\gamma\cdot p$ and $\g\cdot \pp$ and functions (such as $(\gamma\cdot p)^2=p^2$)  thereof~\cite{Naus:1987kv}. \\
 
 \begin{figure}[h]
\centering
\includegraphics[scale=0.20]{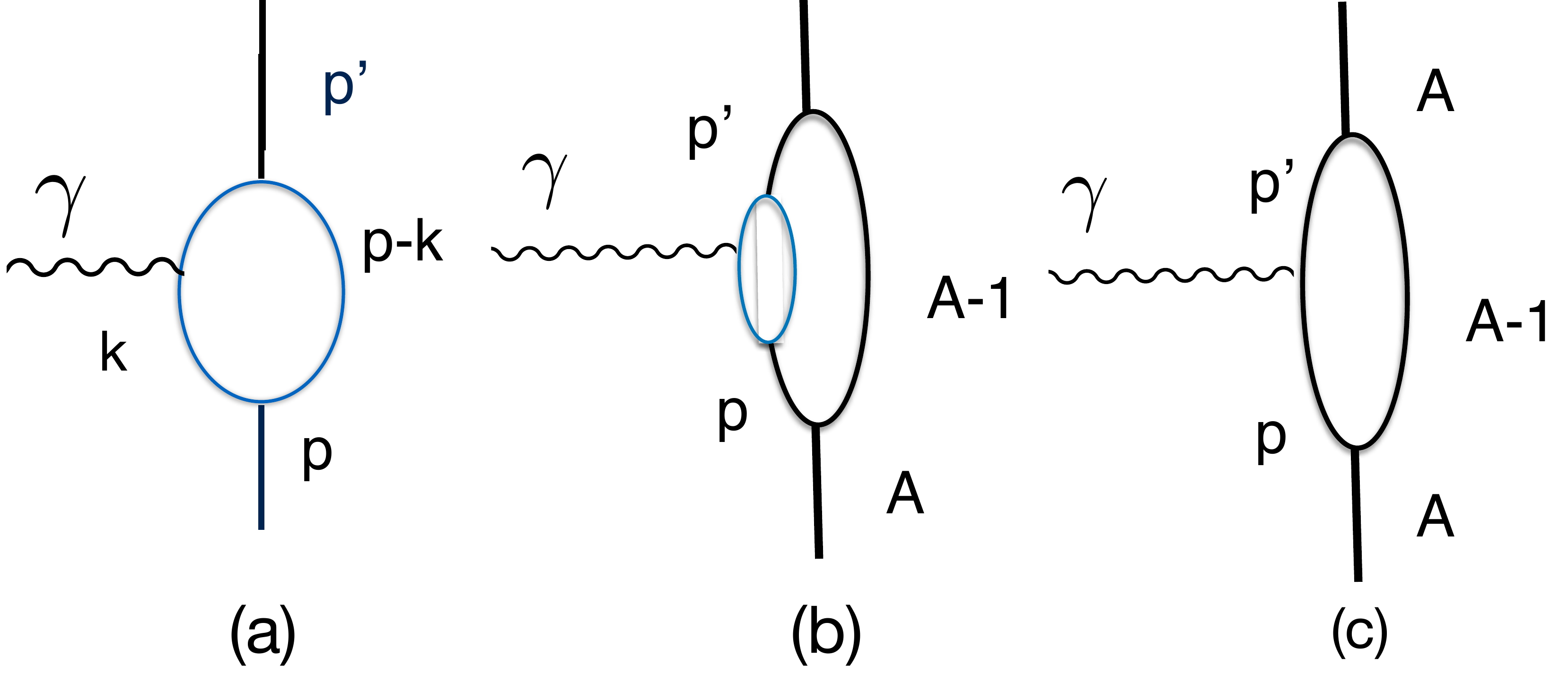}
\caption{\label{rsq}   Photon-nucleon electromagnetic interaction. (a) Photon hits quark in a free nucleon. (b)  Photon hits quark in bound  nucleon.  
c) Photon hits quark in a bare proton bound in the nucleus.}
\end{figure}

As a result  medium modifications of nucleon structure  are determined by the virtuality, $V\equiv p^2-M^2$ and/or  ${p}'^2-M^2$. For elastic scattering on nuclei  Lorentz and time-reversal invariance insures that these   two quantities  are equal. %  The free proton has $V=0$, but this is not so for a bound one. 
The average value of the virtuality can be computed from the spectral function~\cite{CiofidegliAtti:2007ork}, but nuclear wave functions are not presented as a function of specific values of $V$. Therefore  the technical procedure here is  to use a first-order expansion  in powers of $V$.  
At first glance, the fact that the binding energy is small compared to the nucleon mass might seem to make it reasonable to neglect the differences between $p^2,\,\pp^2$  and $M^2$.  However, a better estimate can be obtained from the Schroedinger equation. For example,  within the nuclear Hartree-Fock approximation   a single particle wave function obeys the Schroedinger equation, with a  dominant central binding potential $\cU\,(\ll M)$. Therefore
$\vec{p}^2/(2M) +\cU = -B$, where $B>0$ is the binding energy and   $p^0=M-B$. Then  
$p^2-M^2=(M-B)^2-\vec{p}^2-M^2\approx  2M\cU$. The dependence on the binding energy cancels to first-order in $B$. In the centers of typical nuclei $\cU$ is about  $-50 \,$ MeV~\cite{GFB72}, so that $(p^2-M^2)/M^2 \approx - 0.1$,  significantly different  from zero, but small enough to be considered an expansion parameter.    \\

The detailed examination of the factorization approximation  begins with a study of the  Feynman graphs of  Figs.~1 a,b. The aim is to compute the dependence on the off-mass-shell invariants that appear in the nucleus.  The calculations  are done so that the Ward-Takahashi identity, which  guarantees current conservation, is respected.  For the present models, including the diagram of Fig.~1c along with that of Fig.~1b  is necessary for this to occur.  Furthermore, the models must embody confinement.  These two aspects are dealt with below in connection with arriving at the key result:  
\bea \D F_{1,2}=V{\pd F_{1,2}\over \pd M^2}.\label{key}\eea 
At first glance, this equation seems very odd. How can a property of the proton depend upon its mass,  when it is known to very high precision? The proton  mass can be varied at will in different models.  Moreover,  
results of fundamental lattice QCD calculations of nucleon properties depend implicitly on the proton mass via  quark-mass dependence.  The result \eq{key}  is obtained  when $M^2$ is associated with the four-momentum squared that appears in propagators of  the Bethe-Salpeter  equations determining the  wave functions of the various  models used below.\\%  that is parametrized in terms of the square of the pion mass $m_\pi^2$. In turn the proton mass, $M$ depends on $m_\pi^2$ in those calculations.\\

The effects encapsulated in \eq{key} can be recast in terms of specific two-body currents because the inverse of nucleon propagators contain a  factor of  $V$. The present formulation includes a variety of terms, is very compact and (as to be argued) arises naturally from %simple 
considerations of relativistic dynamics.\\

The next step is to explain how    \eq{key} is derived. Five  different models of the free proton  are used:
\begin{itemize}
\item Quark-diquark, with spin 0 quarks and di-quarks with a scalar vertex function
\item Quark-diquark with spin 1/2 quark, spin 0 di-quark with a scalar vertex function
\item Quark-diquark with spin 1/2 quark, spin 1 di-quark with a vector vertex function (QED)
\item Proton sometimes fluctuates into  its neutron-$\pi^+$ component, with pseudovector coupling.
\item Proton  sometimes fluctuates into  its $\D\pi$ components, with pseudovector coupling.
\end{itemize}
None of these models is  realistic by itself, but each characterizes  a significant aspect of proton structure. \\

%{\Large\color{red}Expand these two paragraphs}
Evaluating the Feynman graph of Fig.~1a for general off-shell kinematics  renders it suitable for inclusion in Fig.~1b. The first-order approximation in $V$ allows the separate study of each term that contributes to medium modifications. The models employed here share   common features, so that  the generality of \eq{key} can be  displayed by discussing only the salient aspects of the models.   
 For each model, the proton wave function involves a vertex function that converts a proton of momentum to a system of two constituents. One of the constituents, denoted by $q$ is charged and interacts with the photon, and the other, denoted by $d$ is a spectator.  This notation is used for both quark-spectator models and pion-spectator models. In each model the three propagators provide a denominator of  the form:
$ D\equiv (k^2-m_q^2) ((k+q)^2-m_q^2)((p-k)^2-m_d^2).\ $ These are combined  with  three Feynman  parameters $x,y,z$ respectively, such that $x+y+z=1$, with a useful symmetry between $x$ and $y$. The factor $D$ can 
thus be re-written:  $D\to(k^2-\Delta)^3$, % and adding and subtracting $M^2$ in two places as:
as 
 \bea \Delta=xy Q^2 +m_q^2(x+y)+z m_d^2 -{p^2+{p'}^2\over 2} z(1-z).\label{doff}\eea
% {\bf mention  pion}
  The on-mass-shell value of $\D$, denoted as  $\Delta_{\rm on}$ is  obtained by replacing  $p^2$ and $\pp^2$ by $M^2$. By adding and subtracting the term 
 $M^2z(1-z) $ one obtains the result that 
 \bea \D= ( 1- V{\partial \over \partial M^2} ) \Delta_{\rm on}.\label{DD}\eea  
% \bea\Delta=xy Q^2 +m_q^2(x+y)+z m_d^2 -M^2z(1-z)} + {Vz(1-z)  = ( 1- V{\partial \over \partial M^2} ) \Delta_{\rm on},\eea  where $\D_{\rm on}$ is $\D$ obtained with $V=0$.
 %The final equality is obtained by realizing the common nature of the dependence on $z(1-z)$. 
 When used in a first-order expansion \eq{DD}  gives one of the terms appearing in \eq{key}.\\
 
The terms in the numerator take  many  forms including: $\psl,\,\psl', p^\m, \,{p'}^\m=(p+q)^\m,\,2k^\m,\, k\cdot k',\,\ksl\ksl',\, (k'\equiv k+q)$ where $\mu$ is the Lorentz-index  of the photon-quark (or photon-pion) vertex.  
Let's start with the term  $\psl$, which is re-written as   to first-order in $V$ as follows:
\bea \psl= %M+(\psl-M)= 
M +{p^2-M^2\over \psl+M}\approx  M+{V\over 2M}=(1+V {\partial \over \partial M^2})M, \label{fo}\eea
 and we see the  pattern emerging. The same manipulations can be done   
  for $\psl'$. Another term that enters is $p^\m$.  Calculations are done in the Breit frame,  with $\m=0$ or in the Drell-Yan frame with $\m=+$.  Then  the identity $2p^\m= \g^\m\psl+\psl'\g^\m+ i \sigma^{\m\nu}q_\nu$ 
is useful because the manipulations for $\psl,\,\psl'$ described above are applicable. The term involving $\s^{\m\nu}$ contributes only to the on-mass-shell part of $F_2$. \\

The models involving struck pions contain a numerator term of the form $2(k^\m+p^\m z-q^\m)\to 2p^\m\,z$ because of parity and the use of either of the two mentioned frames.  The model with an intermediate $\D$ contains terms of the form $k\cdot k' $ and $\ksl\ksl'$. Upon applying  the stated  variable transformations, one finds
\bea k\cdot k'  %\to (k+zp -qy)\cdot(k+zp+(1-y)q)\\
\to k^2+ z^2p\cdot\pp+{Q^2\over2} z(1-z).\eea  The $k^2$ term is evaluated along with the denominators that are discussed above. The third term does not involve off-shell proton kinematics. The term $p\cdot\pp$ may be re-written as $q\cdot p+p^2={1\over 2}(\pp^2+p^2-q^2)$, and subtracting and  adding $2M^2$ leads again to the result of \eq{key}.  The manipulations needed to handle the term $\ksl\ksl'$ are essentially the same, upon using \eq{fo}. % {\bf one more invariant???}
\\

The net result is that \eq{key} emerges from each   term. The general argument      is that for each of the terms that enter one may add and subtract the on-shell expression. To first-order in $V$ all terms in the difference between the on-and off-mass-shell expressions can be expressed as a derivative.\\

The next step is to describe how the 
 Ward-Takahashi (WT)  identity~\cite{Peskin:1995ev} is maintained. If this is respected for all values of $p$ and $\pp$, electron-nucleus interactions will satisfy current conservation.     This identity states that the amplitude $\G^\m(p+q,p)$ for a photon of momentum $q$ to be  absorbed by a fermion of momentum $p$ is related to the fermion-propagator 
$S(\psl)={1\over \psl-M_0-\S(\psl)},$  via
\bea q\cdot \G (p+q,p)=S^{-1}(\psl+\qsl)-S^{-1}(\psl) ,\eea %=\qsl -\S(\psl+\qsl)+\S(\psl),\eea
where  $M_0$ the bare mass and $\S(p)$ the self-energy of the  fermion. (A similar identity is obtained for photon absorption on a charged pion.)  Satisfying the WT identity  is absolutely necessary for high-precision nuclear calculations to be valid. \\ %Satisfying this identity is necessary for any model of the nucleon or nuclear form factor.\\ %\footnote{The present definition of the propagator omits the standard factor of $i$} 

If one evaluates the term of Fig.~1 (a), in which the photon-quark interaction is denoted as $\G^{(q)}$ one finds that
\bea q\cdot \G^{(q)}=\S(\psl)-\S(\psl'),\eea
in which the right-hand-side  vanishes if $p^2=M^2$ and $\pp^2=M^2$. Thus the graph of Fig.~1a is a reasonable model for free protons. However, when the proton is bound in the nucleus, as in Fig.~1b, the momenta $p$ and $\pp$ are off the mass-shell and the WT identity is not respected. This problem is fixed by including the graph of Fig~1c. In that case one obtains
$q\cdot \Gamma =(\slashed{p}+\slashed{q}) -\psl -{\large(\Sigma((p+q)^2)-\Sigma(p^2))\large)}=S^{-1}(p+q)-S^{-1}(p).$ The first two terms arise from Fig.~1c, and the next two from Fig.~1b.\\ %For initial and final on-mass-shell kinematics the difference in self-energies cancels, and current conservation is satisfied.  \\

 The next step is to handle quark-confinement. Detailed evaluations of the Feynman graph of Fig.~1b fail dramatically to obey  the factorization approximation, \eq{FA}, if the quark propagator is taken to be that of a free quark.  To see this, examine \eq{doff}. For on shell kinematics with $p^2=M^2,\,\pp^2=M^2$ the value of $\D$ is positive for all values of $x,y$ and $z$ provided the  stability condition $M<m_q+m_d$ is obeyed. A similar stability condition holds for pion-baryon intermediate states. In evaluating the Feynman diagram of Fig.~1b, one integrates over all values of $p^2$,  so that $\Delta$ can be negative. This means that the in medium proton form factor is complex-valued. The free form factor is real-valued, so the factorization approximation \eq{FA} must break down. Moreover, the singularity associated with lack of confinement makes plays havoc in  numerical integration, and the existence of such singularities in models is unphysical  because nuclei are stable.  Finally, the appearance of zeros in the quantity $\D$    means that an expansion of nucleon properties in terms of the virtuality will not converge because the nuclear wave function  will admit very large values.  \\
 
 The negative value of $\D$ can also be understood by examining the proton self-energy, $\S(\pp^2)$ which involves the denominator  $(k^2-m_q^2)((\pp-k)^2-m_d^2)\to k^2 +\pp^2 u(1-u)-m_q^2(1-u)-m_d^2u,$ where $u$ is another Feynman parameter with $0\le u\le1$. This denominator has zeros for values of $\pp$ such that $\pp^2>(m_q+m_d)^2$. This is also the condition required to knock a quark out of the proton. In  Fig.~2, the final $q$ and $d$ can both be on the mass shell whenever  $\pp^2>(m_q+m_d)^2$.  This feature arises from the lack of including effects of confinement.\\
  \begin{figure}[h]
\centering
\includegraphics[scale=0.30]{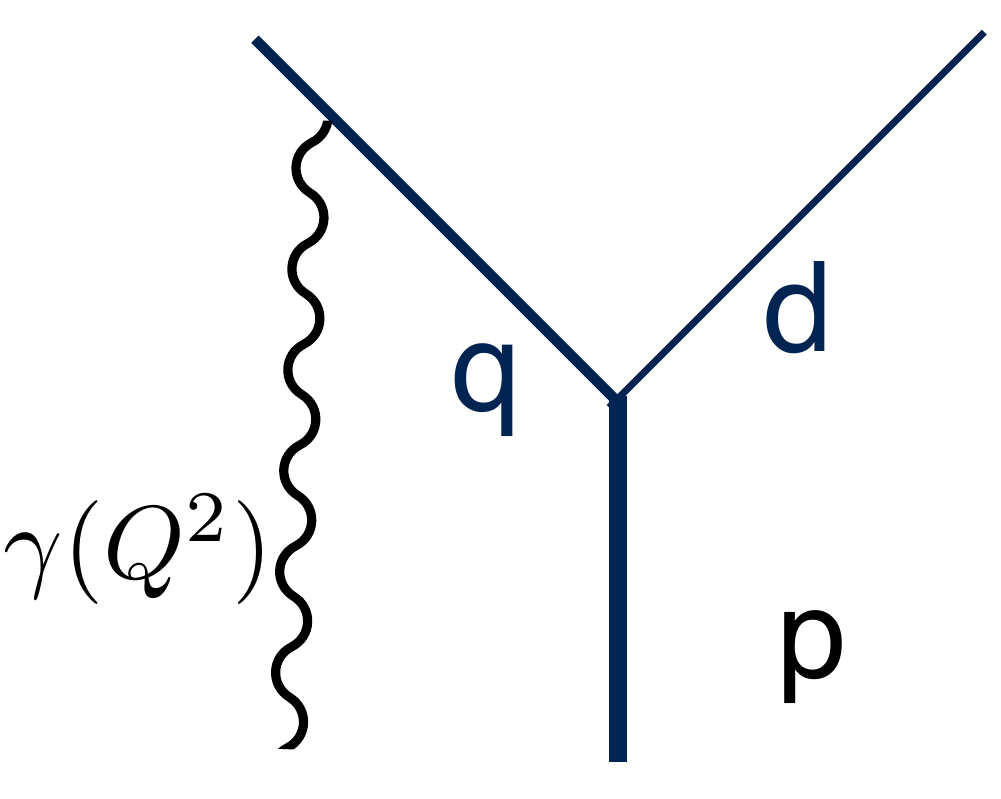}
\caption{\label{rsq1} (Color online)  Photon of momentum $q^2=-Q^2$ hits quark in a free  proton of four-momentum $p$. Final quark  and di-quark can both be on the mass shell.  
 }
\end{figure}

 Some means of implementing the  main feature (no singularities) of confinement must be 
 included in  the present models. This can be done  using an idea suggested by  solutions to Dyson-Schwinger
equations, see the review    \cite{Roberts:1994dr}. The procedure is to  use quark (or di-quark) masses that occur in complex conjugate pairs. 
Complex conjugate singularities in quark propagators have  been studied in  connection with   confinement  in Refs.~\cite{Atkinson:1978tk,Cornwall:1980zw,Munczek:1983dx,Bhagwat:2002tx,Bhagwat:2003vw}, and  % Additional   studies have modeled 
Euclidean space lattice data can be modeled
with propagators that have time-like complex conjugate singularities \cite{Alkofer:2003jj,Alkofer:2003jk}. Ref.~ \cite{Tiburzi:2003ja} used a model with complex conjugate poles to compute parton and generalized parton distributions. \\

The relevance here is that using a di-quark (spectator)  propagator of the form 
\bea S_C(p)=\sum_{\l=-1,1} {1\over p^2+m_d^2+i\l\e}.\label{PV}
\eea
in Euclidean space removes the unphysical singularities.  The previous analysis of the effects of virtuality  has been applied using \eq{PV}  to the models discussed above with the result \eq{key}. Furthermore,   detailed Euclidean space calculations using the models described above have shown that the results of using such a propagator can be obtained in Minkowski space by simply  using a complex di-quark mass and obtaining the form factors by  taking the real part of the computed  amplitude. The net result is that using complex-valued quark masses removes the unphysical singularities  initially present in the simple models used here.  This is necessary to justify  expansions in terms of virtuality.\\

The next step is to apply the key result, \eq{key}. It is worthwhile to study
the proton charge  radius defined \cite{Miller:2018ybm} by
$ {r_E^2}\equiv -6 G_E'(0) $ in this first evaluation. Applying  \eq{key} then leads to a change in $r_E^2$ given by
$   \delta r_E^2 ={V }{ }{\partial r_E^2\over \partial M^2}$
 To evaluate this expression
  it is necessary to know 
how the  proton radius depends on its  mass. This derivative was explicitly computed to be negative for the five models discussed above. Given the negative values of the virtuality, the proton radius must expand when it is  bound in the nuclear medium!\\

It  is useful to take a more general approach by  examining a broader range of models and lattice QCD calculations. For the models presented here  an off-mass-shell proton is equivalent to an on-mass-shell proton of a mass less than $M$. This is because off-shell effects of the energy denominator $\D$ of \eq{doff}  (in which $p^2$ replaces the $M^2$ that appears as an eigenvalue of the Bethe-Salpeter eqaution) and  the off-shell effects of the numerator end up looking like \eq{key}.
Based on obtaining \eq{key} in all of the models, I  assume that it  is a generally valid, first-order treatment of the effects of virtuality and discuss the necessary derivative in  a  broader  framework. \\  % However, one reaches a qualitative conclusion at this point--the radius of the proton increases when it is bound in the nucleus. This is because both the virtuality, $V$ and the partial derivative are negative.\\

Let's first examine  the venerable
MIT bag model~\cite{Chodos:1974pn}
In its most simplistic version (with vanishing quark masses)  the bag radius is inversely proportional to the mass of the nucleon. This translates to the result ${M^2\over r_E^2}{\partial r_E^2\over \partial M^2}=-1.$
 The counterpoint to this model is the non-relativistic quark model as presented in  {\it e.g} Ref.~\cite{Isgur:1979be}, in which  harmonic oscillator confinement is used with the size parameter:
 $b^2\propto {1\over m_q}.$ This leads to  ${M^2\over r_E^2}{\partial r_E^2\over \partial M^2}=-{1\over 2}$ because the $u,d$ quark mass $m_q\approx M/3$.  There are other models in which pionic effects are very important.
 One example is found in the work of \cite{Beg:1973sc} in which the dominant isovector contribution to the square of the nucleon radius is proportional to $\ln{M/m_\pi}$, where $m_\pi $ is the pion mass. This may be translated (using Eq. (4.2) of that paper)  to  $ {M^2\over r_E^2} {\partial r_E^2\over \partial m_\pi^2}=-0.6{M^2\over m_\pi^2}$,  potentially a  very  large effect. More detailed versions of this idea have been used to interpret the pion-mass dependence of the results of lattice QCD calculations,  see {\it e.g.} ~\cite{Hall:2013oga}. If the proton mass increases with increasing pion mass (a reasonable expectation), the derivative ${\pd r_E^2 \over \pd M^2}$ is again negative. \\
 
 In each of  model there is a connection between  the proton mass and radius that occurs via a   fundamental aspect of the model, such as bag radius, quark mass  and the pion mass. 
 The trend of all of these models is clear-  ${\pd r_E^2\over \pd M^2 }<0,$ which, along with the fact that  $V<0$ means that the radius of a  bound proton must be larger than that of a free one.\\

  It is natural to turn to existing  lattice QCD calculations of the proton radius because  hadronic properties  have long been computed as a function of quark masses (as expressed through the pion mass). The mass parameter that sets the mass scale is typically the square of the pion mass. In principle one can determine ${\partial r_E^2\over \partial m_\pi^2}$ and also $\partial M\over \partial m_\pi^2$ from lattice QCD calculations. The ratio of these two quantities gives the desired derivative.  \\

Lattice QCD calculations of the proton charge radius have made significant recent progress~\cite{Constantinou:2014tga,Hasan:2017wwt,Jang:2018djx,Jang:2019jkn,Alexandrou:2017ypw,Alexandrou:2018sjm,Bhattacharya:2013ehc,Ishikawa:2018rew}. Some difficulties involving lattice spacing, finite volume effects,  disconnected graphs and the need to use extrapolations to extract charge radii. Current  lattice results typically undershoot experiment by about 25 \%. %, see {\it e.g.}~\cite{Hasan:2017wwt,Jang:2018djx,Alexandrou:2017ypw,Alexandrou:2018sjm}.  
Many calculations focus on the dominant isovector radius because of the cancellation of disconnected graphs. Another  persistent difficulty has been in making calculations at the small value of the physical pion mass. Ref.~\cite{Alexandrou:2018sjm} did calculations at $m_\pi=135$ MeV, and the recent calculation by the PACS collaboration (parallel array computer system)  performed calculations at a pion mass of 146 MeV, using a large lattice size~\cite{Ishikawa:2018rew}. That work compiled recent results for $r_E^2$.  Another calculation spanning pion masses from 135 to 320 MeV is that of Ref.~\cite{Jang:2018djx}. Ideally one could use the lattice calculations to determine ${\partial r_E^2\over \partial m_\pi^2}$ and ${\partial r_E^2\over \partial M^2}$.  %The size of the present error bars make it difficult to make a precise determination of  ${\partial r_E^2\over \partial m_\pi^2}$. 
Using Fig~13 of Ref.~\cite{Ishikawa:2018rew} allows a determination that 
$
 {\partial r_E\over \partial m_\pi^2}
\approx-3 \,{\rm fm\, GeV^2}
$
with  a {\it large} uncertainty.  This number is obtained by making a linear fit between the geometric mean of the values of the radii at the two lowest mass points~~\cite{Ishikawa:2018rew,Alexandrou:2018sjm} and a precise value at $m_\pi=220 $ MeV~\cite{Bhattacharya:2013ehc}. The more recent lattice data from~\cite{Jang:2018djx,Jang:2019jkn} use an analytic parametrization of the $m_\pi^2$ dependence of their results. Using their formula and taking the isovector result to be dominant gives $
 {\partial r_E\over \partial m_\pi^2}
\approx-2.6 \,\pm 0.3 \,{\rm fm\, GeV^2}
$
%% from MC.nb
 The nucleon mass is well-described as a function of the pion mass as $M\approx M_0+1.14\, {\rm GeV}^{-1}m_\pi^2$
\cite{WalkerLoud:2008bp}. See also ~\cite{Alberg:2012wr}  for an interpretation of that formula. Using the  results~\cite{Jang:2018djx,Jang:2019jkn}  one finds
\bea  \delta r_E = %{V } {\partial r_E\over \partial M^2}=
{V } {\partial r_E\over \partial m_\pi^2}{\partial m_\pi^2\over \partial M^2} =  -{V\over M^2}  (1.1\pm0.1)\,{\rm fm}\label{result}
,
\eea
with the  only source of uncertainty  arising from $ {\partial r_E\over \partial m_\pi^2}$. Taking     $V/M^2=-0.1$  (from its value at  the  nuclear center)  leads to an increase of the proton radius by about 0.11 fm or about 12\%.      The sign is well-determined as the product of two numbers that are each strongly constrained to be negative. The magnitude is less well determined, but is a reasonable estimate. \\

Can this effect be measured? See~\cite{Wang:2018wfz} for a perspective.
A  0.11 fm increase is actually a  rather large effect, but the increase is much  smaller at the edge of the nucleus. %    Attempts to observe a medium modification of the proton  radius are reviewed in~\cite{Hen:2016kwk}.
Tantalizing hints have been seen  a several arenas. For example,  in quasi-elastic electron nucleus  scattering \cite{Strauch:2012nra} and in the Coulomb sum rule (inclusive $(e,e')$ scattering)~\cite{Morgenstern:2001jt,Paolone:2018jup}.   No definitive evidence  for such a change has been seen so far, see {\it e.g.} \cite{Sick:1985ygc}. In high-momentum transfer quasielastic scattering the final-state knocked out proton is essentially free, whereas for  the present case of elastic scattering the initial and final nucleons are equally virtual. Thus the 12\% reported here would be reduced by a factor of two, even before accounting for reductions caused when the reaction occurs near the nuclear surface. Moreover, the quasi-elastic  measurements  have complications  due to  the presence of final state interactions and related issues with current conservation. \\%Obtaining accurate data for the Coulomb sum rule  is complicated by the need to make a longitudinal-transverse separation in electron scattering and other issues. New data are eagerly awaited.\\
%he virtual photon energy up to infinity, which includes the kinematically unavailable time-like region. \\

The  increase in radius seen here represents a violation of the factorization approximation \eq{FA}. To understand its impact let's examine 
%A quantitative study of the accuracy of this approximation  has never been done, so an initial one is provided here for
 the case of A=3. The current status of root-mean-square charge radii is summarized in Table I. 
\begin{table}[ht]
\caption{Charge radii  (in fm) for $^3$H and $^3$He. The first two rows are from experiment, the next two from theory.}  
\centering
\begin{tabular}{c c  c}
\hline\hline
Ref.  & $^3$H   &  $^3$He  \\ [0.5ex] % inserts table %heading
\hline
SACLAY \cite{Amroun:1994qj}  &1.76$\pm$ 0.09& 1.96$\pm$0.03 \\
Bates   \cite{Beck:1987zz} &1.68$\pm$ 0.0347& 1.87$\pm$0.03 \\
GFMC \cite{Pieper:2001ap}  & 1.77$\pm$ 0.0131&1.97$\pm$0.0125   \\
$\chi$EFT \cite{Piarulli:2012bn}  & 1.756$\pm$0.006 & 1.962$\pm$ 0.004 \\
[1ex]
\hline
\end{tabular}
\label{table:Aeuqal3}
\end{table}
The nuclei $^3$H and $^3$He  are related by isospin invariance, which is an approximate symmetry. Coulomb (and other smaller) effects cause $^3$He to be less bound and larger.  The standard procedure, see {\it e.g} \cite{Pieper:2001mp} for  computing the charge radii of 3-body nuclei    is to expand each of the terms of \eq{FA} to first-order in $Q^2$. This results in an expression: $R_A^2=R_{\rm pt}^2+r_E^2$ (if the neutron contribution is neglected). The proton radius $\sim 0.86 $ fm is not small compared to the nuclear charge radii displayed in Table I. The average virtuality for $^3$He is reported in Ref.~\cite{CiofidegliAtti:2007ork} as $V/M^2=-0.073$. Given this number and \eq{result} the resulting shift in the proton radius is about 0.08 fm. This corresponds to   a 2\% increase in the $^3$H charge radius comparable to   the current  experimental uncertainties, but  within reach of present experimental goals. Furthermore the increase of 0.08 fm is  much larger than  that produced by the effects of meson exchange currents or variations in the cutoff of chiral perturbation theory  found in  \cite{Piarulli:2012bn}. Note also that  the difference in radii between $^3$He and $^3$H may be strongly  affected by the effects of virtuality   because a struck proton in $^3$He is influenced by one $pn$ interaction and one $pp$ interaction, while a struck proton 
in $^3$H is influenced by  two $pn$ interactions. The $T=0$ $pn$ interactions are stronger than the $T=1$ $pp$ interactions,  an important feature  in a 
diverse set of reactions~\cite{Hen:2016kwk,Duer:2018sxh,Schmookler:2019nvf}.  The net result is that the increase in the proton radius suggested here is relevant for understanding the properties of nuclei with $A=3$.\\

The calculations performed here are for nuclear medium effects of electromagnetic form factors. There are many general features of these calculations, so that one may speculate that the key result of \eq{key} extends to other matrix elements of other one-body operators, $\cal O$ such that $\langle{\cal O}(p^2)\rangle\approx \langle{\cal O}(M^2)\rangle+(p^2-M^2) {\partial \over \partial M^2}\langle{\cal O}(M^2)\rangle$. The simplicity of this relation is very appealing. If it is valid,    describing  a wide variety of medium effects from the unified viewpoint of examining the dependence on virtuality may be possible.\\

I summarize. The field-theoretic calculations discussed here show that  
 nucleons must be  modified when bound in nuclei. Five  different models yield the result \eq{key}. % F_{1,2}(Q^2)=V{\partial\over \partial M^2}F_{1,2}(Q^2)$}
The necessary derivatives with respect to mass that appear in that equation may  be computed using lattice QCD. Perhaps other proton properties can also be treated this way.
The present work provides a new approach to understanding nuclear modifications of nucleon properties and  strengthens the connection between lattice   
  QCD calculations and nuclear physics. \\
  
  {\bf Acknowledgments}
G.A. Miller would like to thank the Lab for Nuclear Science at MIT, the Southgate Fellowship of Adelaide University (Australia), the Bathsheba de Rothchild Fellowship of Hebrew University (Jerusalem), the Shaoul Fellowship of Tel Aviv University, the Physics Division of Argonne National Laboratory and the U. S. Department of Energy Office of Science, Office of Nuclear Physics under Award Number DE-FG02-97ER-41014 for support that enabled this work.
I also thank P. C. Tandy, G.F. Bertsch,  I. Clo\"et, M. Consantantinou, B. Holsetin, V. Koch,  C. Monahan, C. Roberts,  M.J. Savage, S. Sharpe,  M. Strikman and S. Syritsyn for useful discussions. I thank H-W Lin, R. Gupta  and the members of the PNDME collaboration for providing numerical data in advance of their arXiv posting.

%\section*{Appendix A-Details}


\begin{thebibliography}{99}
 
\bibitem{Sargsian:2002wc} 
  M.~M.~Sargsian {\it et al.},
  %``Hadrons in the nuclear medium,''
  J.\ Phys.\ G {\bf 29},  R1 (2003)
  
 \bibitem{Aubert:1983xm} 
  J.~J.~Aubert {\it et al.} [European Muon Collaboration],
  %``The ratio of the nucleon structure functions $F2_n$ for iron and deuterium,''
  Phys.\ Lett.\  {\bf 123B}, 275 (1983).
 
  %\cite{Arnold:1983mw}
\bibitem{Arnold:1983mw} 
  R.~G.~Arnold {\it et al.},
  %``Measurements of the a-Dependence of Deep Inelastic electron Scattering from Nuclei,''
  Phys.\ Rev.\ Lett.\  {\bf 52}, 727 (1984).
  %doi:10.1103/PhysRevLett.52.727
  \bibitem{Mardor:1990cr} 
  Y.~Mardor {\it et al.},
  %``K+ total cross-sections as a test for nucleon 'swelling.',''
  Phys.\ Rev.\ Lett.\  {\bf 65}, 2110 (1990).
 \bibitem{Hen:2016kwk} 
  O.~Hen, G.~A.~Miller, E.~Piasetzky and L.~B.~Weinstein,
  %``Nucleon-Nucleon Correlations, Short-lived Excitations, and the Quarks Within,''
  Rev.\ Mod.\ Phys.\  {\bf 89},  045002 (2017) 
 
 \bibitem{Cloet:2019mql} 
  I.~C.~Clo\"et {\it et al.},
  %``Exposing Novel Quark and Gluon Effects in Nuclei,''
  arXiv:1902.10572 [nucl-ex]., in press J.\ Phys.\ G.
  \bibitem{Wang:2018wfz} 
  R.~Wang, R.~Dupre, Y.~Huang, B.~Zhang and S.~Niccolai,
  %``Flavor-dependent EMC effect from a nucleon swelling model,''
  Phys.\ Rev.\ C {\bf 99},  035205 (2019)
  \bibitem{Ruiz:2016gne} 
  R.~F.~Garcia Ruiz {\it et al.},
  %``Unexpectedly large charge radii of neutron-rich calcium isotopes,''
  Nature Phys.\  {\bf 12}, 594 (2016)
  
  \bibitem{Antognini:2015moa} 
  A.~Antognini {\it et al.},
  %``Experiments towards resolving the proton charge radius puzzle,''
  EPJ Web Conf.\  {\bf 113}, 01006 (2016)
 % doi:10.1051/epjconf/201611301006
%  [arXiv:1509.03235 [physics.atom-ph]].
  
    \bibitem{12} Jefferson Laboratory experiment 12-14-009, 
  ``Ratio of the electric form factor in the mirror nuclei $^3$He and $^3$H", 
  J. Arrington and D. Higinbotham, Spokespersons, %https://https%3A%2F%2Fwww.jlab.org%2Fexp_prog%2Fproposals%2F14%2FPR12-14-009.pdf&usg=AOvVaw0N8Y79LImEtDXEiXOf-7FP

  \bibitem{Gomez:2017cwj} 
  J.~Gomez,
  %``The Mirror Nuclei$^{3}\hbox {H}$ and$^{3}\hbox {He}$ Program at JLab,''
  Few Body Syst.\  {\bf 58},  97 (2017).


 \bibitem{Amaldi1950} E. Amaldi, G. Fidecaro, and F. Mariani,
 %On the coulomb scattering of $\mu$ mesons by light nuclei,"
  Il Nuovo Cimento, {\bf 7} , 553 (1950).
 \bibitem{Villi1959} C. Villi,   
 % A remark on the interpretation of the electron-nucleus scattering experiments"
 Nucl. Phys. 10, 166 (1959).
 \bibitem{Naus:1987kv} 
  H.~W.~L.~Naus and J.~H.~Koch,
  %``Electromagnetic Interaction of an Off-shell Nucleon,''
  Phys.\ Rev.\ C {\bf 36}, 2459 (1987).
   
  \bibitem{CiofidegliAtti:2007ork} 
  C.~Ciofi degli Atti, L.~L.~Frankfurt, L.~P.~Kaptari and M.~I.~Strikman,
  %``On the dependence of the wave function of a bound nucleon on its momentum and the EMC effect,''
  Phys.\ Rev.\ C {\bf 76}, 055206 (2007)
  \bibitem{GFB72}
  G.~F.~Bertsch,`` The practioner's shell model",
North Holland  Pub. Co. , Amsterdam  (1972)
  
  
    
  \bibitem{Peskin:1995ev} 
  M.~E.~Peskin and D.~V.~Schroeder,
  ``An Introduction to quantum field theory,''
  Addison-Wesley, Menlo Park, 1995
 
\bibitem{Roberts:1994dr} 
  C.~D.~Roberts and A.~G.~Williams,
  %``Dyson-Schwinger equations and their application to hadronic physics,''
  Prog.\ Part.\ Nucl.\ Phys.\  {\bf 33}, 477 (1994)


  %\cite{Atkinson:1978tk}
\bibitem{Atkinson:1978tk} 
  D.~Atkinson and D.~W.~E.~Blatt,
  %``Determination of the Singularities of the Electron Propagator,''
  Nucl.\ Phys.\ B {\bf 151}, 342 (1979).
  
  \bibitem{Cornwall:1980zw} 
  J.~M.~Cornwall,
  %``Confinement and Chiral Symmetry Breakdown: Estimates of f(pi) and of Effective Quark Masses,''
  Phys.\ Rev.\ D {\bf 22}, 1452 (1980).
  

\bibitem{Munczek:1983dx} 
  H.~J.~Munczek and A.~M.~Nemirovsky,
  %``The Ground State q anti-q Mass Spectrum in QCD,''
  Phys.\ Rev.\ D {\bf 28}, 181 (1983).

%  \cite{Bhagwat:2002tx}
\bibitem{Bhagwat:2002tx} 
  M.~Bhagwat, M.~A.~Pichowsky and P.~C.~Tandy,
  %``Confinement phenomenology in the Bethe-Salpeter equation,''
  Phys.\ Rev.\ D {\bf 67}, 054019 (2003)
  %\cite{Bhagwat:2003vw}
\bibitem{Bhagwat:2003vw} 
  M.~S.~Bhagwat, M.~A.~Pichowsky, C.~D.~Roberts and P.~C.~Tandy,
  %``Analysis of a quenched lattice QCD dressed quark propagator,''
  Phys.\ Rev.\ C {\bf 68}, 015203 (2003)
  
  
  
  %\cite{Alkofer:2003jj}
\bibitem{Alkofer:2003jj} 
  R.~Alkofer, W.~Detmold, C.~S.~Fischer and P.~Maris,
  %``Analytic properties of the Landau gauge gluon and quark propagators,''
  Phys.\ Rev.\ D {\bf 70}, 014014 (2004). 

\bibitem{Alkofer:2003jk} 
  R.~Alkofer, W.~Detmold, C.~S.~Fischer and P.~Maris,
  %``Analytic structure of the gluon and quark propagators in Landau gauge QCD,''
  Nucl.\ Phys.\ Proc.\ Suppl.\  {\bf 141}, 122 (2005)
\bibitem{Tiburzi:2003ja} 
  B.~C.~Tiburzi, W.~Detmold and G.~A.~Miller,
  %``Complex conjugate poles and parton distributions,''
  Phys.\ Rev.\ D {\bf 68}, 073002 (2003).


\bibitem{Miller:2018ybm} 
  G.~A.~Miller,
  %``Defining the proton radius: A unified treatment,''
  Phys.\ Rev.\ C {\bf 99},   035202 (2019)
 \bibitem{Chodos:1974pn} 
  A.~Chodos, R.~L.~Jaffe, K.~Johnson and C.~B.~Thorn,
  %``Baryon Structure in the Bag Theory,''
  Phys.\ Rev.\ D {\bf 10}, 2599 (1974).
  \bibitem{Isgur:1979be} 
%  N.~Isgur and G.~Karl,
  %%``Ground State Baryons in a Quark M
  N.~Isgur and G.~Karl,
  %``Ground State Baryons in a Quark Model with Hyperfine Interactions,''
  Phys.\ Rev.\ D {\bf 20}, 1191 (1979).
  
  
  
  
  
 \bibitem{Beg:1973sc} 
  M.~A.~B.~Beg and A.~Zepeda,
  %``Pion radius and isovector nucleon radii in the limit of small pion mass,''
  Phys.\ Rev.\ D {\bf 6}, 2912 (1972).
  
  \bibitem{Hall:2013oga} 
  J.~M.~M.~Hall, D.~B.~Leinweber and R.~D.~Young,
  %``Chiral extrapolations for nucleon electric charge radii,''
  Phys.\ Rev.\ D {\bf 88},   014504 (2013)
   
 

   \bibitem{Alexandrou:2013joa} 
  C.~Alexandrou, M.~Constantinou, S.~Dinter, V.~Drach, K.~Jansen, C.~Kallidonis and G.~Koutsou,
  %``Nucleon form factors and moments of generalized parton distributions using $N_f=2+1+1$ twisted mass fermions,''
  Phys.\ Rev.\ D {\bf 88},   014509 (2013)
   \bibitem{Jang:2018djx} 
  Y.~C.~Jang, T.~Bhattacharya, R.~Gupta, H.~W.~Lin and B.~Yoon,
  %``Updates on Nucleon Form Factors from Clover-on-HISQ Lattice Formulation,''
  PoS LATTICE {\bf 2018}, 123 (2018)
  \bibitem{Jang:2019jkn} 
  Y.~C.~Jang, R.~Gupta, H.~W.~Lin, B.~Yoon and T.~Bhattacharya,
  %``Nucleon Electromagnetic Form Factors in the Continuum Limit from 2+1+1-flavor Lattice QCD,''
  arXiv:1906.07217 [hep-lat].

  \bibitem{Hasan:2017wwt} 
  N.~Hasan, J.~Green, S.~Meinel, M.~Engelhardt, S.~Krieg, J.~Negele, A.~Pochinsky and S.~Syritsyn,
  %``Computing the nucleon charge and axial radii directly at $Q^2=0$ in lattice QCD,''
  Phys.\ Rev.\ D {\bf 97},   034504 (2018)
  %\cite{Alexandrou:2017ypw}
\bibitem{Alexandrou:2017ypw} 
  C.~Alexandrou, M.~Constantinou, K.~Hadjiyiannakou, K.~Jansen, C.~Kallidonis, G.~Koutsou and A.~Vaquero Aviles-Casco,
  %``Nucleon electromagnetic form factors using lattice simulations at the physical point,''
  Phys.\ Rev.\ D {\bf 96},  034503 (2017)
 \bibitem{Alexandrou:2018sjm} 
  C.~Alexandrou, S.~Bacchio, M.~Constantinou, J.~Finkenrath, K.~Hadjiyiannakou, K.~Jansen, G.~Koutsou and A.~V.~A.~Casco,
  %``Proton and neutron electromagnetic form factors from lattice QCD,''
  arXiv:1812.10311 [hep-lat].

  \bibitem{Ishikawa:2018rew} 
  K.~I.~Ishikawa {\it et al.} [PACS Collaboration],
  %``Nucleon form factors on a large volume lattice near the physical point in 2+1 flavor QCD,''
  Phys.\ Rev.\ D {\bf 98},   074510 (2018)
  
  

  \bibitem{Bhattacharya:2013ehc} 
  T.~Bhattacharya, S.~D.~Cohen, R.~Gupta, A.~Joseph, H.~W.~Lin and B.~Yoon,
  %``Nucleon Charges and Electromagnetic Form Factors from 2+1+1-Flavor Lattice QCD,''
  Phys.\ Rev.\ D {\bf 89},  094502 (2014)

%  \cite{Constantinou:2014tga}
\bibitem{Constantinou:2014tga} 
  M.~Constantinou,
  %``Hadron Structure,''
  PoS LATTICE {\bf 2014}, 001 (2015)
  
  

  %\cite{Alexandrou:2018sjm}
%\cite{Alexandrou:2011db}
\bibitem{Alexandrou:2011db} 
  C.~Alexandrou {\it et al.},
  %``Nucleon electromagnetic form factors in twisted mass lattice QCD,''
  Phys.\ Rev.\ D {\bf 83}, 094502 (2011)

  

     
  \bibitem{WalkerLoud:2008bp} 
  A.~Walker-Loud {\it et al.},
  %``Light hadron spectroscopy using domain wall valence quarks on an Asqtad sea,''
  Phys.\ Rev.\ D {\bf 79}, 054502 (2009)
  \bibitem{Alberg:2012wr} 
  M.~Alberg and G.~A.~Miller,
  %``Taming the Pion Cloud of the Nucleon,''
  Phys.\ Rev.\ Lett.\  {\bf 108}, 172001 (2012)
  \bibitem{Strauch:2012nra} 
  S.~Strauch,
  %``Hadron medium modifications,''
  EPJ Web Conf.\  {\bf 36}, 00016 (2012).
  \bibitem{Morgenstern:2001jt} 
  J.~Morgenstern and Z.~E.~Meziani,
  %``Is the Coulomb sum rule violated in nuclei?,''
  Phys.\ Lett.\ B {\bf 515}, 269 (2001)
  

  \bibitem{Paolone:2018jup} 
  M.~Paolone [Jefferson Lab Hall-A E05-110 Collaboration],
  %``Measuring the Coulomb sum rule at JLab,''
  AIP Conf.\ Proc.\  {\bf 1970},   020010 (2018).
  \bibitem{Sick:1985ygc} 
  I.~Sick,
  %``On the size of nucleons in the nuclear medium,''
  Phys.\ Lett.\  {\bf 157B}, 13 (1985).
  \bibitem{Amroun:1994qj} 
  A.~Amroun {\it et al.},
  %``H-3 and He-3 electromagnetic form-factors,''
  Nucl.\ Phys.\ A {\bf 579}, 596 (1994).
  
\bibitem{Beck:1987zz} 
  D.~Beck {\it et al.},
  %``Isoscalar and isovector form factors of H-3 and He-3 for Q below 2.9 fm-1 from electron-scattering measurements,''
  Phys.\ Rev.\ Lett.\  {\bf 59}, 1537 (1987).
  
   \bibitem{Pieper:2001ap} 
  S.~C.~Pieper, V.~R.~Pandharipande, R.~B.~Wiringa and J.~Carlson,
  %``Realistic models of pion exchange three nucleon interactions,''
  Phys.\ Rev.\ C {\bf 64}, 014001 (2001)
  
   \bibitem{Piarulli:2012bn} 
  M.~Piarulli, L.~Girlanda, L.~E.~Marcucci, S.~Pastore, R.~Schiavilla and M.~Viviani,
  %``Electromagnetic structure of A = 2 and 3 nuclei in chiral effective field theory,''
  Phys.\ Rev.\ C {\bf 87},   014006 (2013)

  %\cite{Pieper:2001mp}
\bibitem{Pieper:2001mp} 
  S.~C.~Pieper and R.~B.~Wiringa,
  %``Quantum Monte Carlo calculations of light nuclei,''
  Ann.\ Rev.\ Nucl.\ Part.\ Sci.\  {\bf 51}, 53 (2001)

%\cite{Duer:2018sxh}
\bibitem{Duer:2018sxh} 
  M.~Duer {\it et al.} [CLAS Collaboration],
  %``Direct Observation of Proton-Neutron Short-Range Correlation Dominance in Heavy Nuclei,''
  Phys.\ Rev.\ Lett.\  {\bf 122},  17, 172502 (2019)
\bibitem{Schmookler:2019nvf} 
  B.~Schmookler {\it et al.} [CLAS Collaboration],
  %``Modified structure of protons and neutrons in correlated pairs,''
  Nature {\bf 566},    354 (2019).


 \end{thebibliography}
\end{document}